# On the remote coherence of polariton condensates in 1D microcavities: a photoluminescence study

M. D. Martín[1,2], E. Rozas[1,2], C. Antón[1,2,†], P. G. Savvidis[3,4,5], and L. Viña[1,2,6]

[1] Departamento de Física de Materiales, Universidad Autónoma de Madrid, 28049 Madrid, Spain
[2] Instituto de Ciencia de Materiales "Nicolás Cabrera", Universidad Autónoma de Madrid, 28049 Madrid, Spain
[3] Westlake University, 18 Shilongshan Rd, Hangzhou 310024, Zhejiang, Peoples R China
[4] Westlake Institute of Advanced Study, Institute of Natural Sciences, 18 Shilongshan Rd, Hangzhou 310024, Zhejiang, Peoples Republic of China
[5] Department of Nanophotonics and Metamaterials, ITMO University, 197101 St. Petersburg, Russia
[6] Instituto de Física de la Materia Condensada, Universidad Autónoma de Madrid, 28049 Madrid, Spain

**Introduction**

Semiconductors materials are key elements in the present communication and information era. They are important not only for their support on the technological side, but also because they offer convenient platforms in which different aspects of fundamental physics can be studied. In particular, when semiconductors are illuminated with light of the appropriate energy, a promotion of an electron from the valence band to the conduction band occurs, leaving a hole in the valence band. The Coulomb interaction between these two charges leads to the formation of an exciton, the solid-state quantum-mechanical equivalent of an atom [1].

Since the implementation of Molecular Beam Epitaxy reactors, a great progress has been made in the growth of semiconductors, particularly in the production of very thin layers of a semiconductor with a given energy gap sandwiched between thicker layers of a wider gap material. These structures, called quantum wells (QWs), allow the confinement of excitons inside of them and bring forward quantum confinement effects, such as binding energy and oscillator strength enhancement, together with an energy shift of the exciton emission energy [2, 3]. Furthermore, the refinement of epitaxial growth techniques allows to place these QWs inside optical resonators, typically in the shape of a Fabry-Perot cavity. These cavities, built with distributed Bragg reflectors (DBRs), made of alternating layers of two different semiconductor materials, can reach extremely large quality factors and efficiently confine a photon between the top and bottom DBRs [4]. Tuning the energy of both, confined exciton and photon, by means, for example, of adjusting the QW width and the cavity length, opens the way to the achievement of strong radiation-matter interaction, provided the decay rates of both oscillators (exciton and photon) are smaller than the frequency difference between them. If this situation is met, the exciton "dresses" with the photon and becomes a polariton. Under these conditions, the photon is reversibly absorbed to create an exciton, which gives back the photon when it annihilates. Additionally, when the exciton and the photon have exactly the same energy the strong radiation-matter interaction lifts their energy degeneracy, giving rise to the upper (UPB) and lower polariton branches (LPB), separated by the so-called Rabi splitting [5, 6]. Such a strong coupling regime can be easily achieved in semiconductor microcavities, where the first experimental evidence of a strong coupling was published in 1992 [4]. It is worth mentioning, that the exciton dressing occurs also in the case of weak radiation-matter interaction, but the emission of a photon by the annihilation of the exciton is then an irreversible process.

After Weisbuch *et al.*'s pioneering work [4], the years that followed witnessed the blooming of microcavity research, rendering at first the linear properties of polaritons: the dispersion relation was measured by means of angle-resolved spectroscopy [6], the tuning of the cavity energy, exploiting a wedge purposely introduced in its thickness, was demonstrated together with the tuning of the exciton energy, varying the lattice temperature [7] or applying an external electric field [8], and eventually the polariton linear dynamics was resolved [9]. Polariton nonlinear properties was an effervescent research topic for many years. The bottleneck in the polariton relaxation [10] hindered the observation of indubitable polariton nonlinearities for many years, but ultimately this field granted access to the observation of many physical phenomena, restricted previously to the atomic physics field, such as Bose-Einstein condensation (BEC) or superfluid propagation, but at much higher temperatures than those required in atomic physics. The properties of an exciton-polariton BEC differ from those of other known condensates. This is due to the short lifetime of polaritons, which provides the great advantage of studying the BEC phase and coherence directly, just by collecting the photons leaking out of the cavity. In 2006 the first experimental demonstration that gathered all the appropriate evidences of polariton BEC was published [11], shortly followed by the demonstration of superfluid motion [12, 13], vortex persistence [14, 15] and long-lasting coherence [16] as well as large spatial extension [17] of a polariton condensate. All of these milestones benefited from the ease with which polaritons can be created and detected by external optical means. Excellent reviews on the properties of polaritons have been published in the last years [18, 19]. More recently, the enhancement of the exciton binding energy found in transition metal dichalcogenides and lead halide perovskites, has led to the achievement of polariton condensation [20-22] and lasing [23] at room temperature. The superfluidity of polariton condensates has also been recently revisited [24] and discussed instead as the propagation of a fluid with low viscosity.

Upon the discovery of the outstanding nonlinear behavior of polaritons and their ease of use, a new research field arose in the pursue of employing polaritons in new concept devices: polariton interferometers [25], logic gates [26, 27] or transistors [28 - 30] have been recently reported. Even zero dimensional polaritons, confined in microcavity pillars, have been produced [31 - 33], and two dimensional micropillar lattices are now employed to emulate graphene and its remarkable properties [34]. All of these polariton devices are very precisely designed using sophisticated lithographic techniques that guarantee the polariton confinement not only along the MBE growth direction but also along perpendicular directions, leaving, for example, only a well-defined longitudinal channel for the movement of polariton condensates [27, 28, 35, 36]. Such one-dimensional (1D) microcavity structures are the focus of this review paper, particularly the remote coherence of 1D polariton condensates that have never been in physical contact. In 1984 Anderson [37] discussed this long-standing issue of quantum mechanics and Pitaevskii and Stringari followed up in 1999 with their proposal to address the problem by measuring the interference fringes in momentum space [38]. Early evidences of remote coherence of polariton condensates in k-space can be found in the literature [27, 39] though they were a byproduct and could only be hinted on the time-resolved photoluminescence (PL) data, as shown in Figure 1 (extracted from [27]). Panel (b- 3) displays the time evolution in real space of two polariton condensates moving along a 1D structure and controlled using two excitation beams. The condensates bounce back and forth between the beams (0-50 μm) or between one of the beams and the edge of the structure (50-100 μm). It is possible to see interferences in the vicinity of each of the bounces as a result of the spatial overlap of forward and reflected polaritons. Additionally, panel (b-iii) displays the concurrent time-resolved emission but in momentum space. Interference fringes can be seen between -1 and +1 $\mu m^{-1}$ as the two condensates move left/rightwards at the same time and with the same speed; their wavevectors coincide, giving rise to the observed interferences.

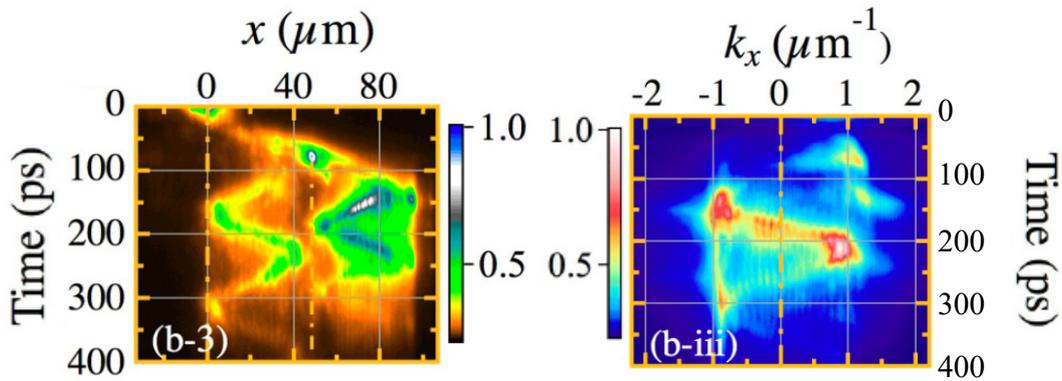

**Figure 1.-** Taken from [27]. Panel (b-3/b-iii) displays the time evolution of the space-/momentum-resolved PL of polariton condensates propagating in a 1D waveguide. The emission intensity in both panels is coded in a logarithmic, normalized, false color scale.

Coherence in real space has been amply studied in cold atoms [40-42], excitons [43, 44] and polariton condensates [11, 17, 45 - 48]. In recent years, novel time-resolved experiments have been performed to tackle the phase-locking mechanisms of polariton condensates [49-51]. Christmann *et al*. addressed the phase locking dynamics of two independent, spatially separated and expanding condensates by means of time-resolved and interferometric measurements, under pulsed, non-resonant excitation [50]. Ohadi *et al*. studied the dissipative coupling between two spatially separated condensates leading to a relative in-phase or out-of-phase locking between them [51].

In this manuscript we will gather clear experimental evidences of remote coherence between two polariton condensate droplets that have never overlapped in real space and discuss how these interferences in momentum space can be used to estimate the critical temperature for the BEC like transition.

## Samples and experimental details

Even though our experimental findings have been collected using several samples, with slight differences between them, the main scheme of our 1D microcavities, shown in Figure 2 (a), is the following: it consists of a two-dimensional high-quality (Q factor ~ 16000) $5\lambda/2$ AlGaAs-based microcavity, where four sets of three 10 nm GaAs QWs have been inserted at the antinodes of the electromagnetic field. The strong radiation-matter coupling results in a 9.2 meV Rabi splitting. Then this planar microcavity sample is patterned through reactive ion etching to obtain linear waveguides with dimensions 20 x 300 $\mu m^2$. The pattern, consisting of ridges (bound by yellow rectangles) and pillars of various sizes, has been repeatedly sculpted over the sample, as shown in Figure 2 (b). Further details about the samples can be found in Ref 52.

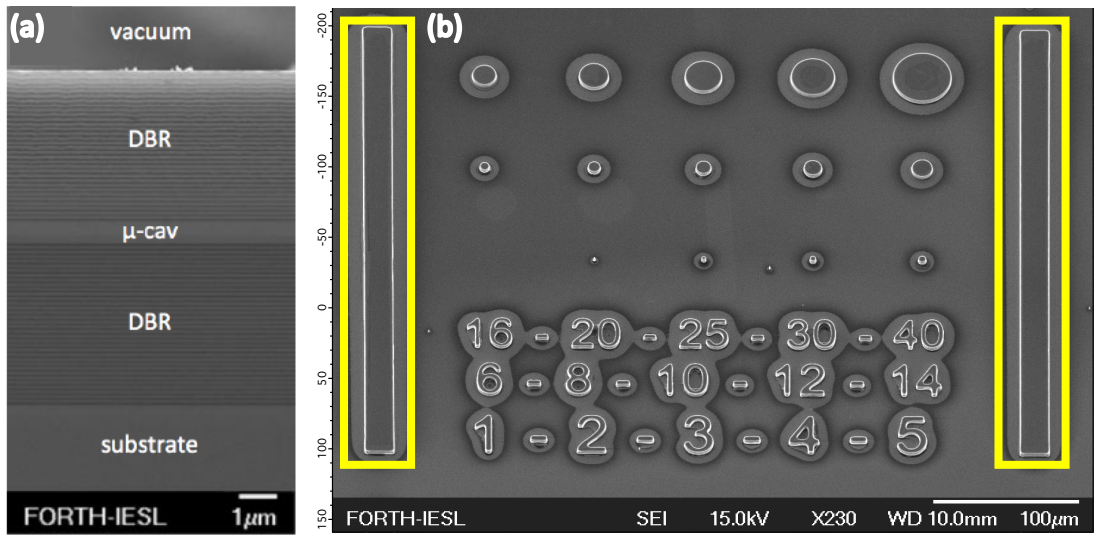

**Figure 2.-** (a) SEM image of the microcavity structure along the growth (vertical) direction. (b) Top-view SEM image of the unitary pattern sculpted over the sample by reactive ion etching. The yellow rectangles highlight the ridge structures used in our experiments.

The samples are kept in a cold finger cryostat where their temperature can be controlled and varied from 10 to 50 K. They are photoexcited with 2 ps long pulses derived from a Ti:Al$_2$O$_3$ laser. In our experiments, either one or two laser pulses are used to excite the sample. The beams are focused using a microscope objective (NA = 0.4, f = 10 mm) and are precisely controlled to have the same power density, to impinge on the sample perpendicularly to its surface and to be separated by 70 μm on the sample surface. The excitation energy is tuned to the exciton levels, a few meV above the LPBs. The polariton emission is collected by the same microscope objective and focused on the slit of an imaging spectrometer using a long focal length lens. The spectrometer filters the energy at which the experimental data are recorded, being able to register only the emission related with the phenomena that we want to study. At the exit of the spectrometer we have installed a charged coupled device (CCD) for time-integrated measurements, and a streak camera for time-resolved ones. An example of the real space, time-integrated emission can be found in Figure 3, which displays on panel (a) the real space emission map (x vs y) obtained after far non-resonant excitation at the first minimum above the DBRs' stop band. The emission covers the whole length of the structure, insinuated by the dashed line, due to the polariton condensate's expulsion from the excitation area (at x = 0) [34]. There is however no energy selected, so the exciton cloud (immobile, at x = 0), the travelling polaritons (moving, in opposite directions, away from x = 0) and two localized states at the edges of the structure (static, at x ~ ± 150 μm) are seen in the same image. The picture changes drastically when the energy is resolved [Figure 3 (b)], so that those different emitting species can be distinguished. The exciton cloud appears between 1.546 and 1.543 eV, the polariton condensates travel with an energy of 1.5405 eV, and below this energy, the two localized states appear at 1.5395 eV. Looking at this image it is easier to understand how the polariton condensates acquire a velocity: at the excitation area, around x = 0, they possess a large potential energy that is going to be reduced by changing into kinetic energy, so they move away from x = 0 with a constant total energy. The potential energy, and therefore the speed, can be controlled tweaking the pumping power, as a larger power yields a greater LPB blue-shift and hence a higher potential energy and polariton speed.

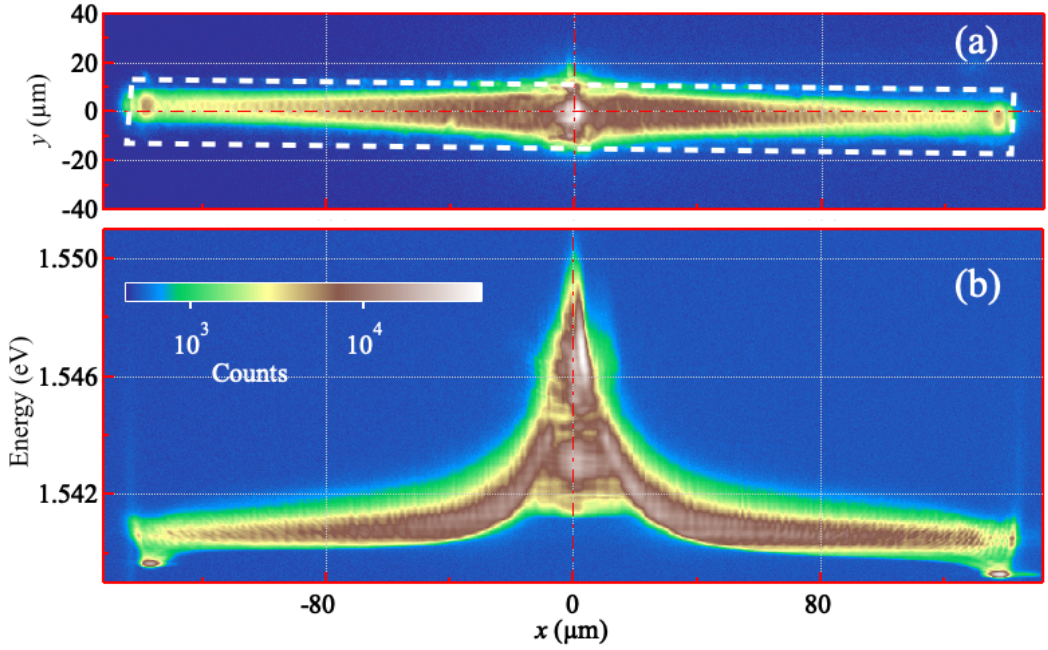

**Figure 3.-** (a) Real space map of a 1D microcavity pumped at x = 0 (white area). The dashed line denotes the geometry of the structure. (b) Energy relaxation process along the structure, integrating across y. The excitation laser is well above 1.55 eV. Polariton condensates are expelled from the excitation area in opposite directions and travel with an energy of 1.5405 eV. Below this energy, two localized states are observed at the edges of the structure, at 1.5395 eV. Both images have been time-integrated and obtained at 10 K and for an excitation power density of 10 kW/cm$^2$.

The introduction of an additional lens in the PL setup allows us to measure the momentum-space emission maps instead of those of real space. The microscope objective focuses the momentum (angular) distribution of the emission on its back focal plane, commonly known as Fourier plane. The image formed in this Fourier plane, which coincides with the front focal plane of the real-space imaging lens, is collected by this lens while keeping its back focal plane at the appropriate distance from the spectrometer slit. Like this, the momentum distribution is collimated and later focused on the spectrometer slit by the additional lens (dubbed as k-space lens). Further details about the lens arrangements can be found in Ref. 53. Figure 4 (a) displays a standard two-dimensional map of the emission, measured in momentum space ($k_x$ vs $k_y$). The map has been obtained detecting all the energy spectrum of the polariton emission but filtering, using a 800 nm long pass filter, to avoid the saturation of the detectors by the excitation laser and to dim the stronger exciton emission, so it reveals several features: there appear two white (i.e. bright) spots at ($k_x$ ~ ± 1 μm$^{-1}$, $k_y$ ~ 0) corresponding to the polariton condensates moving away from the excitation area. The condensate droplets cannot move along the transverse (y) direction, only along the longitudinal (x) one, and so, their wavevector only has a non-zero $k_x$ component. The image also displays a weaker intensity around $k_x = k_y = 0$ originating from both, the exciton cloud created by the excitation at x = 0 and the polariton condensates trapped at the edges of the structure. Finally, a very faint intensity disc can be seen on the background, related to the emission of excitons inside a rather wide solid angle (and with a higher energy). When this emission map is resolved in energy, we obtain the exciton-polariton dispersion relation [Figure 4 (b)]. It is worth mentioning that in this case, only one of the wavevector components can be detected. We have selected $k_x$ since it is aligned with the free movement direction (i.e. x). The same long pass filter used in Figure 4 (a) is placed before the spectrometer's entrance slit. We obtain the weak exciton emission between 1.546 and 1.543 eV, below, around 1.5405 eV the polariton condensate droplets, travelling with $k_x \sim \pm 1$ µm$^{-1}$ and further down, at 1.5395 eV, the two localized states at the edges of the 1D microcavity are seen.

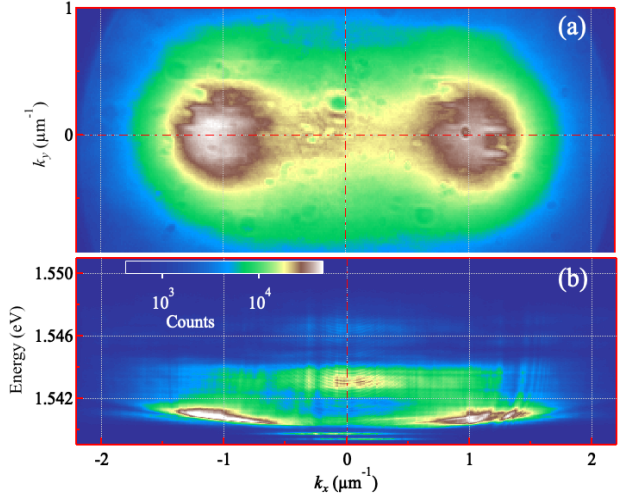

**Figure 4.-** (a) Momentum space map ($k_x$ vs $k_y$) of the emission of a 1D microcavity. (b) Energy dispersion relation. Both images have been time-integrated and obtained at 10 K and for an excitation power density of 10 kW/cm$^2$.

The experiments we shall discuss in the next section are performed in a similar fashion as those presented here. In order to study the remote coherence of distant polariton condensates it is necessary to create at least two condensates, well separated from each other. We have shown above that, after photoexcitation, two polariton condensate droplets are expelled from the excitation area in opposite directions, but these two droplets derive from the same excitation event, so they will for sure be coherent. The creation of two independent condensates requires the use of two excitation beams. The direct consequence of this will be the creation of four droplets, two of them will move to the left of the structure, the other two to the right. We will study the coherence of these pairs of droplets, travelling in the same direction but originating from different excitation positions, in momentum space while the excitation beams, arriving to the sample at the same time, are separated by a constant distance of 70 µm. After demonstrating the remote coherence of polariton condensates we will discuss how we have exploited this coherence and its vanishing with increasing the lattice temperature to estimate the critical temperature for Bose-Einstein condensation.

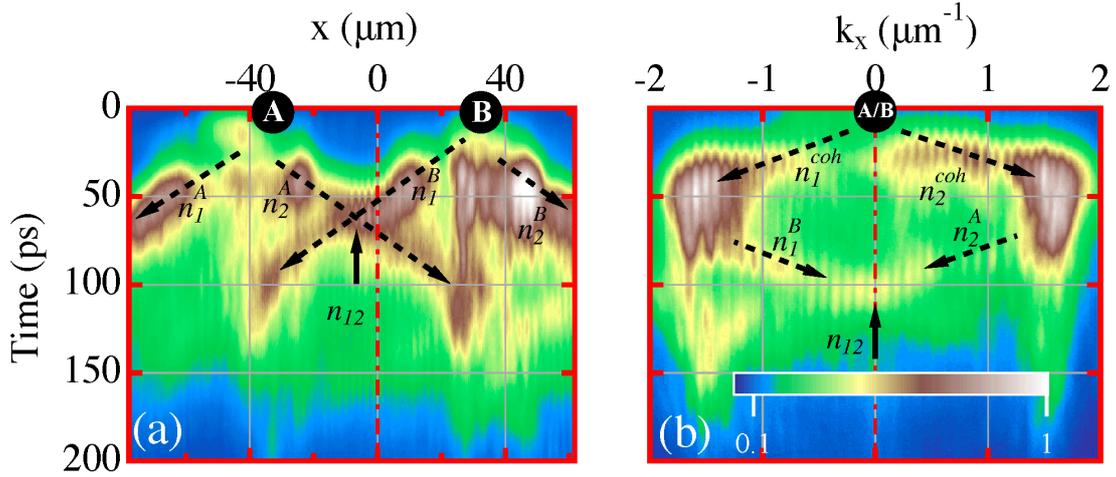

**Figure 5.-** (a) Time-resolved real-space emission map of the traveling polariton condensates. A and B denote the positions of the excitation beams, at ± 35 µm. $n_{1,2}^{A/B}$ denotes the polariton condensate traveling along the 1D microcavity. The subscript (1/2) refers to the condensates traveling to the left/right and the superscript A/B denotes the beam that created the condensate. (b) Time-resolved momentum-space emission map of the traveling polariton condensates. Both excitation beams, A and B, impinge on the sample at normal incidence and therefore appear at $k_x$ = 0 µm$^{-1}$. After that the condensates accelerate and acquire a finite wavevector ± 1.6 µm$^{-1}$. The power density of each excitation beam is 28 kW/cm$^2$. The intensity is coded in a normalized, logarithmic false color scale.

**Experimental results**

Let us start by describing how is the movement dynamics of the four polariton condensate droplets by having a look at it in real space [Figure 5 (a)]. The excitation pulses, A and B, arrive to the 1D microcavity structure at the same time and perpendicularly to the sample surface. They are separated 70 µm, so they appear at x = ± 35 µm, are tuned quasi in resonance with the exciton levels (1.545 eV) and have a power density of 28 kW/cm$^2$. The figure displays the emission intensity map obtained for an energy of 1.5405 eV (that of the traveling polariton condensates) as a function of both the longitudinal coordinate (x) and time. The trajectory of four polariton condensates droplets, labelled $n_1^A, n_1^B, n_2^A$ and $n_2^B$, can be seen. $n$ stands for the polariton density of each droplet, obtained from the square modulus of the polariton wavefunction. The superscripts A and B refer to the excitation beam originating the droplet, and the subscripts 1 and 2 distinguish between the left and right traveling droplets, respectively. Each pair of droplets ($n_1^{A/B}$ and $n_2^{A/B}$) are separated by 70 µm. This distance remains constant as long as the droplets move in the original direction, as confirmed by the parallel slopes of their emission traces. At approximately 66 ps, the droplet traveling to the right from spot A ($n_2^A$) and that traveling to the left from spot B ($n_1^B$) overlap in the vicinity of x = 0. We have highlighted this overlap by adding an arrow pointing towards the crossing area, labelling it as $n_{12}$. Their coexistence leads to the appearance of interference fringes, even if they last only for about 20 ps, as the droplets do not stop moving but go ahead in their respective trajectory until they reach the vicinity of the neighboring excitation spot. They start losing some of their kinetic energy as they approach the potential barrier created by the long-living excitons around the excitation area. Since the do not have enough energy to overcome this barrier, they slowly come to a halt and then regain their kinetic energy as they elastically bounce against the excitation potential. It should be mentioned that there seems to be fringes all through the image, even when there are no overlapping condensates. These fringes do not have the same origin as those we are referring to here as they are related to backscattered polaritons (to be discussed later).

All the peculiarities of this dynamics are confirmed by the time- and momentum-resolved emission map [Figure 5 (b)], measured immediately after acquiring the real-space map, thus ensuring the same excitation conditions. The acceleration of the four droplets, two of them traveling left ($n_1^{coh}$), the other two to the right ($n_2^{coh}$), is observed between 0 and ~ 40 ps, reaching a wavevector $k_x = \pm 1.6$ µm$^{-1}$ from their initial rest ($k_x$= 0). Clear interference fringes are obtained around the maximum values of $k_x$, from ~ 40 to ~ 75 ps, evidencing the remote coherence of two polariton condensate droplets, arising from different and distant excitation areas, that have never been in contact and just move with the same velocity and in the same direction. These are not the only interference patterns observed in Figure 5 (b). A second set of fringes, underlined by a black arrow and labelled $n_{12}$, appears as a result of the interference between the two droplets that have stopped next to their neighboring excitation area. Both emissions occur at $k_x$= 0, and even though the droplets are far from each other in real space, they overlap in momentum space, generating the observed fringes. This overlap in momentum space is preceded by a deceleration of droplets $n_1^B$ and $n_2^A$, as they approach the excitation areas, confirming the dynamics in real space discussed above. Further details about the droplets' full dynamics and a profound discussion about the origin of the constant phase difference between remote droplets can be found in Ref. 54.

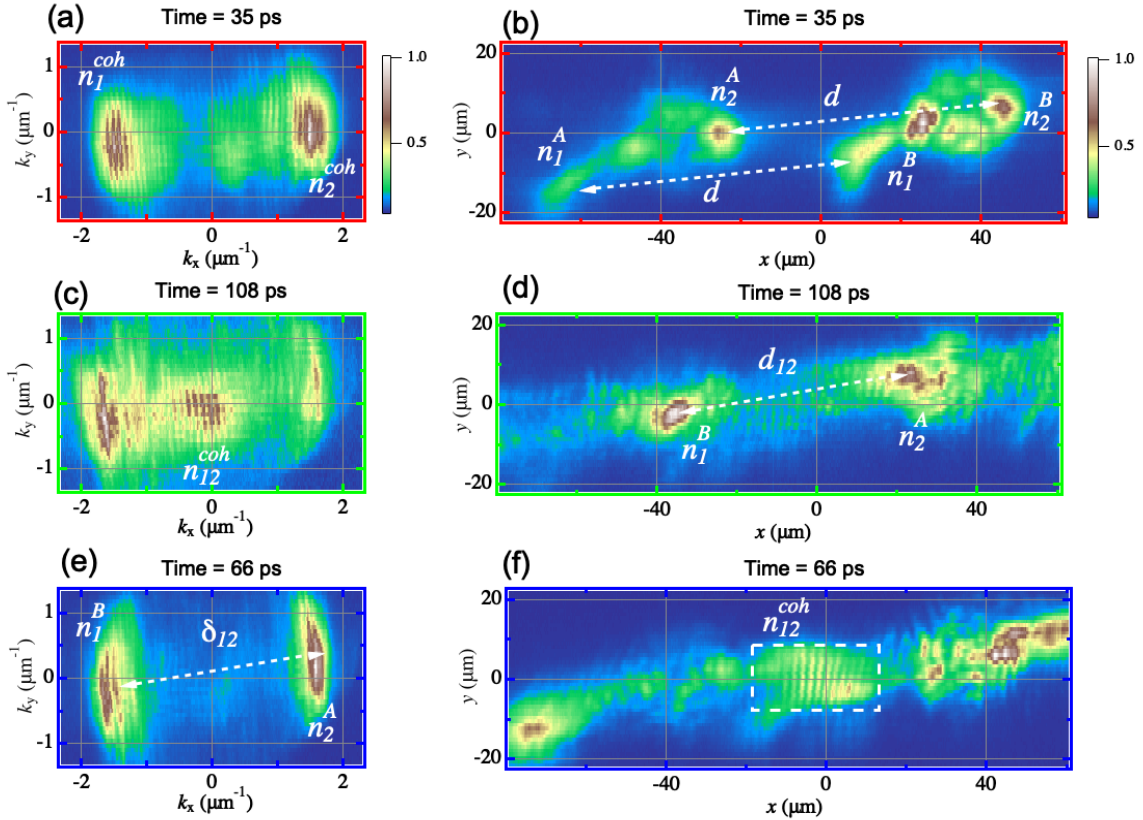

**Figure 6.-** (a) 2D momentum space emission map, 35 ps after excitation. The two polariton condensate droplets, $n_1^{coh}$ and $n_2^{coh}$, can be seen at $k_x = \pm 1.6$ µm$^{-1}$, $k_y$~0, respectively. (b) Corresponding 2D real space emission map, obtained 35 ps after excitation, pinpointing in real space the k-interfering droplets. (c) 2D momentum space emission map 108 ps after excitation, showing interference fringes at $k_x = 0$ as a result of the overlap in k space of the emissions of the bullets halted next to the exciton barriers. (d) Corresponding 2D emission map, at 108 ps delay, locating the k-overlapping droplets. (e) 2D momentum space emission map at 66 ps time delay. (f) Corresponding real space emission map at 66 ps, showing the overlap at $x$~0 of two polariton condensate droplets ($n_2^A$ and $n_1^B$).

Let us have a closer look at the interference fringes in both momentum- and real-space. We will focus our attention on the 2D emission maps collected in Figure 6 for certain, relevant time delays, concentrating on the link between the periodicity of the fringes in each space and the distance between the interfering droplets in the complementary one. In Figure 6 (a) we can see the emission of the droplets in momentum-space 35 ps after the excitation. The droplets traveling to the left/right, labelled as $n_1^{coh}$ and $n_2^{coh}$ respectively, appear at $k_x = \mp 1.6$ µm$^{-1}$. In both spots we find a separation between the fringes of $\Delta k_x = 0.088(5)$ µm$^{-1}$. Applying the fundamental properties of Fourier optics [55], we can relate momentum- and real-space by means of a Fourier transformation and obtain a relation between the periodicity in momentum-space and the distance in real-space. Like this, we can calculate the corresponding distance in real-space (d) between the interfering species given by $\Delta k_x = {2\pi}/{d}$, which in our case yields d = 71(4) µm. Looking at the positions of the interfering droplets in real space [Figure 6 (b)], we find that they are approximately 70 µm apart, in excellent agreement with our findings. It is important to remember that the two interfering droplets do not overlap in real space (in fact they have never been in touch with one another) but in momentum space, moving in the same direction and with the same speed. Similar findings are obtained 108 ps after excitation [Figure 6 (c)], when two of the droplets are stopped in the vicinity of the excitation areas, after crossing each other earlier. The fringes are clearly observed around $k_x \sim 0$, highlighted as $n_{12}^{coh}$, and their separation is $\Delta k_x = 0.108(5)$ µm. The calculated distance of the interfering species is $d_{12}$= 60(4) µm. A close look at the position of the droplets in real space for the same time delay [Figure 6 (d)] reveals that, indeed the distance between the condensates is now approximately 60 µm, as their kinetic energy is not large enough to overcome the potential barrier created by the excitons, immobile in the excitation area, so they stop short before x = ± 35 µm and remain for ~ 20 ps at this shorter (< 70 µm) distance. Finally, for the sake of completeness, let us discuss a biproduct of our experiments, summarized in Figures 6 (e) and (f). Since the Fourier transform link between real and momentum space works both ways, a periodic set of fringes in real space will be correlated with a distance in momentum space (a speed difference) of the interfering parts. At 66 ps two droplets overlap in real space, at x ~ 0 [$n_{12}^{coh}$ in Fig. 6 (f)], giving rise to a set of interference fringes, separated by $\Delta x = 1.99(17)$ µm. According to the relation between real and momentum space, this separation would correspond to a $\Delta \kappa = 3.2(2)$ µm$^{-1}$, where $\Delta \kappa = {2\pi}/{\Delta x}$. Looking at the 2D emission map measured in k space for the same time delay [Fig. 6 (e)], we indeed find that the droplets move with $k_x = \pm 1.6$ µm$^{-1}$, rendering a distance (in k space) of 3.2 µm$^{-1}$, in excellent agreement with that given by the periodicity of the fringes in real space.

Let us discuss now the omnipresent interference fringes observed in the real space emission [Figure 5 (a)]. In the following we will show that they result from the coexistence of forward moving polaritons and backscattered ones. The backscattering takes place in unintended defects and sample irregularities on the 1D microcavity structure. In order to have a better understanding of these interferences we have repeated our experiments using only one excitation beam, arriving at x = 0. Like this the only fringes that we will eventually observe are due to the interaction with backscattered polaritons and not to a more sophisticated polariton interaction. Also, the simplification of the experimental setup allows to easily implement filtering optics for both, real and momentum space, so the images can be properly *cleaned*. Our main findings are summarized in Figure 7. Panel (a) displays the time-resolved real space emission of the traveling condensates, two droplets ejected out of the excitation area, at x = 0, in opposite directions (red and black arrows, respectively) with a speed of 1 µm/ps. There is no filtering applied in this image. Conspicuous, vertical interference fringes (with a period Δx = 2.4 µm) evidence, as we shall prove in the following, the spatial overlap of counter-propagating condensates (forward and backscattered polaritons). The corresponding momentum space, Figure 7 (e), displays a constant |$k_x$| value of ~ 1.3 µm$^{-1}$. Figure 7 (b) displays the time-resolved emission from which the emission originating from x < 0 has been filtered out, so only the droplet moving to the right is observed. Consequently, in momentum space, no emission in the negative wavevector area is registered [Figure 7 (f)], as the droplet moving to the left does not reach the detector. The backscattering of polaritons moving to the right is still taking place, as indicated by the white dashed arrows in Figure 7 (b). In the image displayed in Figure 7 (c) an additional filter in momentum space is applied to the time-resolved emission (already filtered in real-space, removing x < 0). Limiting the detected wavevector range to $0.8 < k_x < 1.6$ µm$^{-1}$ and excluding the contribution of all the other wavevectors from the image [Figure 7 (g)], in particular that of the negative wavevectors, results in the disappearance of the interference with the backscattered polaritons. The time-resolved emission appears clean of fringes. For the sake of completeness, we have also measured the time-resolved emission due to polaritons propagating with negative wavevectors after removing those moving to the left, so that only the $-1.6 < k_x < -0.8$ µm$^{-1}$ range is detected. Like this, only the contribution to the emission arising from backscattered polaritons is measured and it is displayed in Figures 7 (d) and (h) in real- and momentum-space, respectively. In both panels the intensity needs to be multiplied by a factor 10 to appear in the same scale as the rest of the images and the white dashed arrows indicate the contribution of the backscattered polaritons. The same spatial filter can be applied to rid the two-dimensional emission maps of the interference with backscattered polaritons. Figure 8 summarizes the results obtained after sequentially filtering the emission in real and in momentum space, thus eliminating the fringes due to the backscattering. For the sake of conciseness, we will only show the effect of these filters on the real space emission maps. Figure 8 (a) displays the unfiltered real space polariton emission map, obtained 40 ps after excitation with a single laser beam at x = 0. In Figure 8 (b) the fringes on the right-moving polaritons are still seen, even after applying a spatial filter in real space that blocks the polariton droplet moving to the left, i.e. x < 0. The observed fringes must originate from the only polaritons moving to the left that will reach the detector, i.e. those which are backscattered. Blocking the trajectory of those polaritons, employing an additional filter in momentum-space that prevents polaritons with negative wavevectors from reaching the detector (removing $k_x < 0.8$ μm$^{-1}$), the droplet moving to the right appears clean of fringes [Figure 8 (c)]. If the momentum-space filter is applied to allow only the arrival of the backscattered polaritons instead, blocking both $x < 0$ and $k_x > -0.8$ μm$^{-1}$, [Figure 8 (d)], we find that even though their emission is at least two orders of magnitude lower than that of forward moving polaritons, their impact on the dynamics is remarkable, as they are responsible for the appearance of interference fringes all along the duration of the emission. A possible way to diminish the impact of these fringes and eventually suppress the polariton backscattering may be to increase the excitation power, as discussed in Ref. 56.

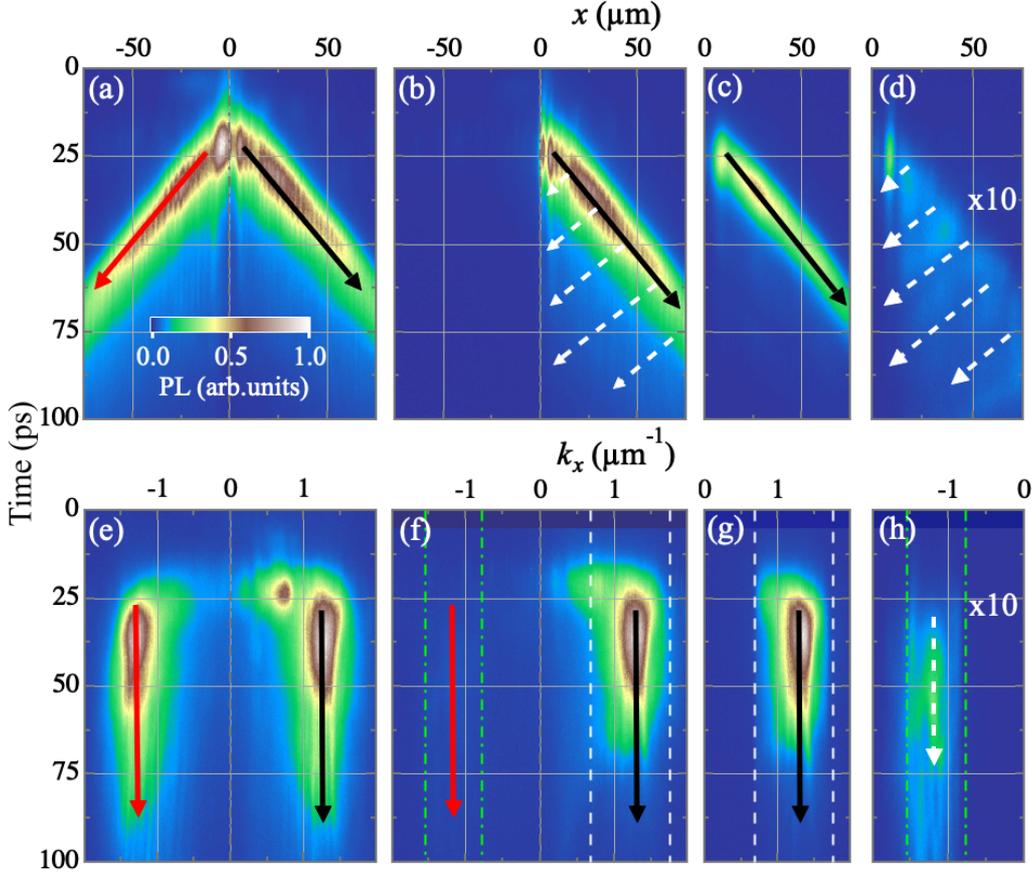

**Figure 7.-** Time-resolved real-/momentum-space distribution of the polariton droplets emission under the following filtering conditions: (a)/(e) no-filtering; (b)/(f) filtered real-space for x < 0; (c)/(g) filtered real-space for x < 0 and 1.6 < $k_x$ < 0.8 µm$^{-1}$ in momentum-space; (d)/(h) filtered real-space for x < 0 and − 1.6 > $k_x$ > − 0.8 µm$^{-1}$ in momentum-space. Polariton emission in panels (d,h) is multiplied by a factor 10. Black (white, dashed) arrows depict the incident (backscattered) polariton population. In panel (f), green, dot-dashed (white, dashed) lines illustrate the region resolved in momentum space in panel (g) [(h)]. A red arrow shows the leftwards moving polaritons. Emission intensity is coded in a linear, normalized, false color scale.

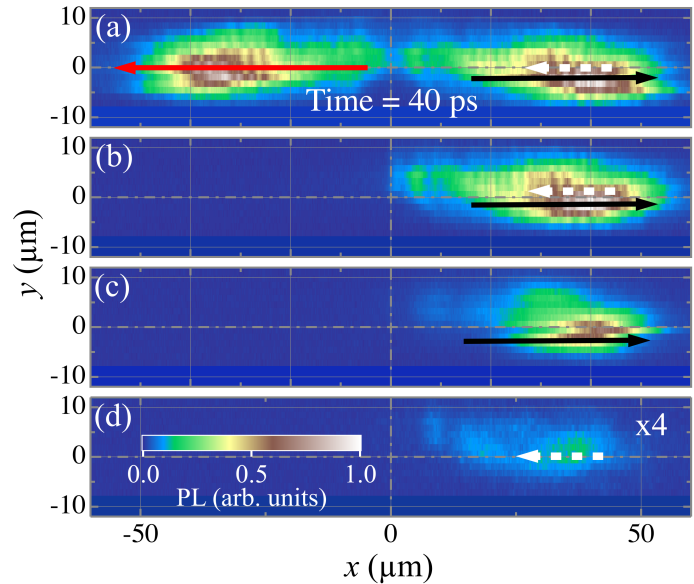

**Figure 8.-** Real-space distribution of the polariton emission, 40 ps after the pulsed excitation at x = 0, under the following filtering conditions: (a) no-filtering; (b) filter in real-space for x < 0; (c) filtered real-space for x < 0 and $k_x < 0.8$ µm$^{-1}$ in momentum-space; (d) filtered real-space for x < 0 and $k_x > -0.8$ µm$^{-1}$ in momentum-space. Polariton emission in panel (d) is multiplied by a factor 4. A red/black arrow depicts polaritons moving left/right and a dashed-line white arrow represents the backscattered polaritons. Intensity is coded in a linear, normalized, false color scale.

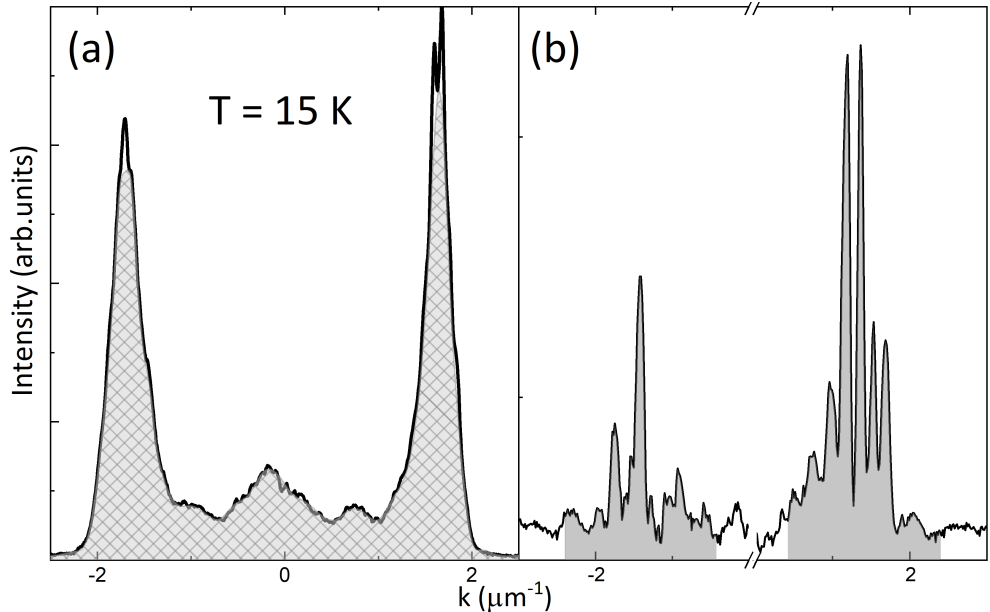

**Figure 9.-** (a) PL emission measured for a lattice temperature of 15 K and integrated during $t_1$, in momentum space. The grey shaded area marks the base line, subtracted from the PL profile (black line), which weighs the non-condensed polariton population. (b) Interference profile after subtraction of the base line shown in (a). The contribution of the condensed polariton population is computed as the light grey area.

Now that we have addressed the remote coherence in momentum space of two polariton condensates that have never been in touch with one another and discussed the effect of backscattered polaritons on the emission, in the following we will focus on how can we use the interference fringes to test the thermal robustness of the polariton condensates and to estimate a critical temperature for the phase transition. The fringes will be somewhat like a thermometer, as they will disappear when the critical temperature is reached and the coherent polariton population vanishes. To do this, we have varied the lattice temperature, linked somehow to that of the condensates, until the interference fringes disappear and concentrated our analysis on three relevant time intervals, which we have already examined: the time during which the droplets move with a constant speed ($t_1$), overlapping in pairs in momentum space, the time in which droplets moving in opposite directions meet in real space ($t_2$) and finally the time interval when the droplets are stopped in the vicinity of the excitation areas ($t_3$), coinciding around $k_x \sim 0$. All three time windows are chosen wide enough to collect the emission contribution relevant for each of them and to assure that the intensity of the fringes remains constant. We integrate the emission in each time range and obtain a profile, which contains the contributions of both condensed and non-condensed polaritons. The profile obtained for $t_1$ at 15 K is displayed in Figure 9 (a). We remove the contribution of the population of the thermal, non-condensed polaritons subtracting a handmade base line from the profile (shaded grey area), leaving much clearer interference fringes that reflect only the contribution of the coherent polariton population [Figure 9 (b)]. Integrating the area underneath the fringes and the base line we can calculate the condensed and the non-condensed polariton populations, respectively, obtaining the condensed fraction $f_C$ as the ratio between the two. We perform a similar analysis for all the lattice temperatures used in our experiments, obtaining $f_C$ as a function of the temperature. Our main findings are summarized in Figure 10.

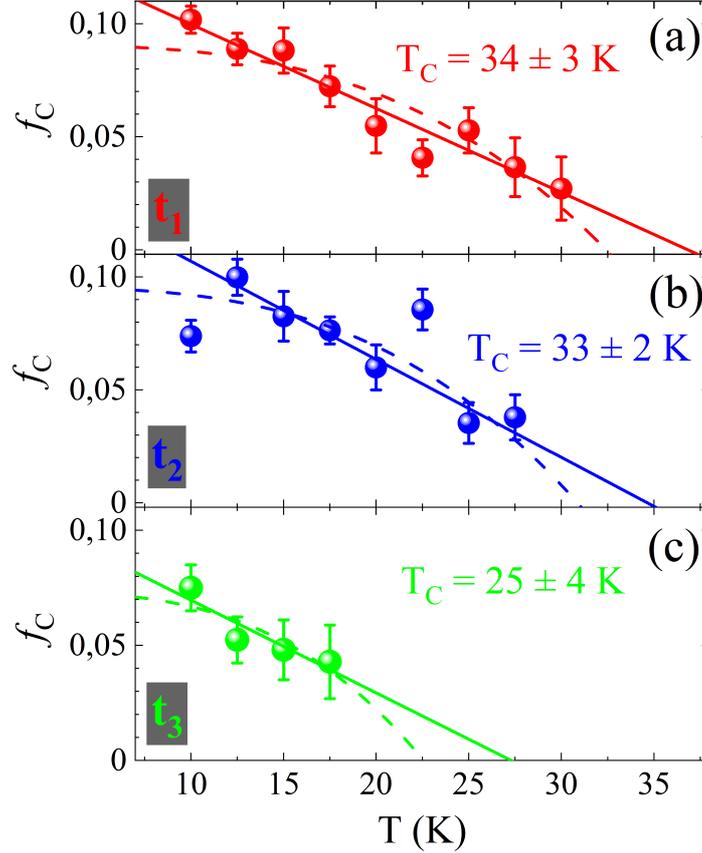

**Figure 10.-** Temperature dependence of the fraction of condensed to non-condensed polariton populations, $f_C$, for $t_1$ (a) in momentum space, $t_2$ (b) in real space, and $t_3$ (c) again in momentum space. The lines in each graph represent a fit to $f_C = f_0 \left[ 1 - \left( T/T_C \right)^\beta \right]$, with $\beta = 1$ (solid line) and $\beta = 3$ (dashed line). Adapted from Ref. 57.

One can see in Figure 10 that $f_C$ decreases with the lattice temperature for all the time intervals considered, evidencing the predominance of the non-condensed population and the vanishing of the condensed one for higher temperatures. That is why we can estimate the critical temperature ($T_C$) when $f_C$ goes to zero. In the literature, there are but two theoretical approaches that address $T_C$ for BEC in atomic systems considering the coexistence of condensed and non-condensed particles. The first theoretical approach is based on a mean field description of a 2D weakly interacting atom gas and predicts a linear reduction with temperature of the condensate fraction [58]. The second one considers a 3D gas of interacting, cold atoms, confined in a cylindrical trap, and predicts a cubic decrease with T of $f_C$ [59]. Based on these theoretical models, we have used $f_C = f_0 \left[ 1 - \left( T/T_C \right)^\beta \right]$ to fit our experimental results, using $\beta = 1$ or $\beta = 3$ to check which one agrees better with our findings. The linear/cubic fit appears as a solid/dashed line on Figure 10.

Even though we have measured the PL for T ≤ 50 K, the worsening of the signal to noise ratio with increasing temperature hinders the reliable determination of the condensate fraction for T > 30 K in $t_1$ and $t_2$, and for T > 20 K in $t_3$. A careful look at the results obtained for $t_1$ [Figure 10 (a)], reveals that about 10% of the polaritons are condensed at the lowest temperature (10 K), and shows that both models give a very similar $T_C$ of 34(3) K. The fact that we cannot definitely assert which theoretical approach is more suitable to fit our results might be related with the non-equilibrium nature of our condensates, since both models are developed for equilibrium BEC. We find very similar results for $t_2$ [Figure 10 (b)], with an initial condensate fraction of approximately 10% and $T_C$ = 33(2) K. The similarity between the experimental findings for $t_1$ and $t_2$ arises from the fact that one ($t_2$) comes right after the other ($t_1$), so the condensate dynamics remains more or less the same and there is no big difference in the condensate fractions of the two time intervals. In contrast, the results obtained for $t_3$ are considerably different: we find an 8% condensate fraction at 10 K and $T_C$ = 25(4) K. These lower values found for both $f_C$ and $T_C$, are directly related with the fact that, for this time interval, the interference (in momentum space, around k ~ 0) occurs when the droplets are in close proximity to the excitonic reservoirs. These reservoirs act as decoherent agents by means of exciton-polariton scattering, rendering a smaller condensate fraction and a higher sensitivity to changes in the temperature. Further details about the thermal robustness of the remote coherence of distant polariton condensates can be found in Ref. 60.

**Conclusions**

In this article we have reviewed the state of the art of the coherence of microcavity-polariton condensates and discussed additional experimental proofs of the remote coherence of two of these condensates that have no knowledge whatsoever of each other's existence. The ease with which polariton condensates can be created and manipulated, together with the symmetry of our excitation scheme, helped us to positively answer to Anderson's long-standing quantum mechanics question. We have also discussed the impact of backscattered polaritons in the observation and analysis of interference fringes, removing them appropriately by filtering in real- and/or momentum-space. Finally, we have used these interferences in reciprocal space between remote condensates to estimate the critical temperature for the Bose-Einstein-like phase transition occurring in semiconductor microcavities.

**Acknowledgements**

We acknowledge the financial support of the Spanish MINECO Grant MAT2017-83722-R. E.R. acknowledges financial support from a Spanish FPI Scholarship No. BES-2015-074708. We thank C. Tejedor for fruitful discussions. P. G. Savvidis acknowledges financial support from the Westlake University foundation and Russian Science Foundation Grant No. 19-72-20120.